\documentclass[aps,prd,showpacs,nofootinbib]{revtex4-1}
\usepackage{latexsym}
\usepackage{amsmath,amsfonts}
\usepackage{amsbsy}
\usepackage{mathrsfs}
\usepackage{color}
\usepackage{psfrag}
\usepackage{enumerate}
\usepackage{amsmath,amssymb,calc,amsfonts}
\usepackage{latexsym}
\DeclareFontFamily{U}{rsfs}{}         
\DeclareFontShape{U}{rsfs}{m}{n}{<5> rsfs5 <6><7> rsfs7          %
  <8><9><10><10.95><12><14.4><17.28><20.74><24.88> rsfs10}{}     %
\DeclareMathAlphabet{\mathfs}{U}{rsfs}{m}{n}                     %
\definecolor{indiagreen}{rgb}{0.07, 0.53, 0.03}
\def\beq{\begin{eqnarray}}
\def\eeq{\end{eqnarray}}

\def\nn{\nonumber\\}

\def\={\stackrel{\Delta}{=}}

\def\lie{\pounds}

\def\half{{\textstyle{\frac{1}{2}}}}

\begin{document}

\title{Quasi-local conformal Killing horizons: Classical phase space and the 
first law}

\author{Ayan Chatterjee}\email{ayan.theory@gmail.com}
\affiliation{Department of Physics and Astronomical Science, Central University 
of Himachal Pradesh, Dharamshala -176215, India.}
\author{Avirup Ghosh}\email{avirup.ghosh@saha.ac.in}
\affiliation{Theory Division, Saha Institute of Nuclear Physics, 1/AF Bidhan 
Nagar, Kolkata 700064, INDIA.}

\begin{abstract}
In realistic situations, black hole spacetimes do not admit a global timlike 
Killing vector field. However, it is possible 
to describe the horizon in a quasi-local setting by introducing the notion of
a quasi-local boundary with certain properties which mimic the properties of a 
black hole inner boundary. Isolated horzons and Killing
horizons are examples of such kind. In this paper, we construct such a boundary of 
spacetime which is null and admits
a conformal Killing vector field. Furthermore we construct the space of 
solutions (in general relativity)
which admits such quasi-local conformal Killing boundaries. We also establish a 
form of first law for these
quasi-local horizons. 
\end{abstract}


\maketitle
\section{Introduction}

A black hole is described to be a region of spacetime where the gravitational
attraction is high enough to prevent even light from escaping to infinity.
In asymptotically flat spacetimes, the impossibility of light escaping to
future null infinity form the appropriate characterization of a black hole. In
other words, this region lies outside the causal past of the future null
infinity $\mathscr{I}^{+}$. The boundary of such a region is called
the event horizon $\mathscr{H}$ \cite{Hawking:1973uf, Wald:1984rg}. To be more 
precise, consider a strongly asymptotically predictable spacetime 
($\mathcal{M}, g_{ab}$). The spacetime is said to contain a black hole if
$\cal{M}$ is not contained in $J^{-}(\mathscr{I}^{+})$. The black hole region 
is
denoted by $\mathscr{B}=\mathcal{M}-J^{-}(\mathscr{I}^{+})$ and the event
horizon is the boundary of $\mathscr{B}$ (alternatively it may also be defined as the future boundary of past of future null infinity: 
$\mathscr{H}=\partial[\,J^{-}(\mathscr{I}^{+})\,]$).
The definition of event horizon thus requires that we are able to construct the
future null infinity $\mathscr{I}^{+}$. This implies that the entire future of
the spacetime needs to be known beforehand to ensure the existence of an event
horizon. Indeed, the condition of strong asymptotic predictibility of spacetime
signifies that we have a complete knowledge of the future evolution. From the 
above consideration, it is clear that $\mathscr{H}$ is a global concept
and it becomes difficult to proceed much further using this definition.
However, the notions simplify for stationary spacetimes which are expected
states of black holes in equilibrium. In equilibrium, these spacetimes admit
Killing symmetries and thus exhibit a variety of interesting features. Indeed,
the strong rigidity theorem implies that the event horizon of a stationary
black hole is a Killing horizon \cite{Hawking:1971vc}. However not all Killing horizons are event horizons. Killing horizons only require a timelike Killing vector field in the neighbourhood of the horizon whereas construction of a stationary event horizon requires a global timelike Killing vector field.

The identification of the event horizon of
a stationary black hole to a Killing horizon was useful to prove the laws
of mechanics for event horizons \cite{Bardeen:1973gs}. It was shown
that in general relativity, the surface gravity $\kappa_{H}$ of a stationary black 
hole must be a constant over
the event horizon. The  first law of black hole mechanics refers to 
stationary space-times admitting an event horizon and small perturbations about 
them. This law states that
the differences in mass $M$, area $A$ and angular momentum $J$ to two nearby 
stationary black hole solutions are related 
through $\delta M=\kappa_{H} \delta A/8\pi + \Omega_{H}\delta J.$ One gets additional 
terms like charge if matter fields are present.
Hawking's proof that due to quantum particle creation, black holes radiate to 
infinity, particles of all species at a temperature $\kappa_{H}/2\pi$,
implied that laws of black hole mechanics are the laws of thermodynamics of 
black holes \cite{Hawking:1974sw}. Moreover, the entropy of the black holes must
be proportional to it's area \cite{Bekenstein:1973ur, Bekenstein:1974ax}.

However, it was realised very soon that this identification of entropy to area leads to new 
difficulties. Classical general
relativity gives rise to infinite number of degrees of freedom but it is not 
clear if the laws of thermodynamics can arise
out of a statistical mechanical treatment of these classical information (see 
\cite{Wald:1995yp}). 
One must find ways to extract quantum degrees of freedom of general relativity. 
The framework of Killing Horizon was broadened to understand the origin of 
entropy and black hole thermodynamics \cite{Wald:1993nt, Iyer:1994ys, 
Jacobson:1993vj, Youm:1997hw, Carlip:1999cy, Dreyer:2013noa, Ghosh:2014pha}.  
It turned out that the framework of Isolated Horizons (IH) was more suited to address these questions from the perspective of loop quantum gravity 
\cite{Ashtekar:1998sp, Ashtekar:2000sz, Ashtekar:2000hw, Ashtekar:2001is, 
Ashtekar:2001jb, Chatterjee:2006vy, Chatterjee:2008if}. It is argued that 
the effective quantum degrees of freedom which capture the thermodynamic 
information of black holes are localised, more precisely, reside on the 
horizon. Isolated horizons are suited for this description since they capture 
only
the local information; isolated horizons are local descriptions of horizons and 
unlike event horizons, do not require 
the global history of spacetime \cite{Smolin:1995vq, Krasnov:1996tb, 
Rovelli:1996dv, Ashtekar:1997yu, Ashtekar:1999wa, Ghosh:2006ph, Ghosh:2008jc, 
Ghosh:2011fc, 
Ghosh:2013iwa}. It arises that the effective field theory induced on 
a IH is a Chern- Simons theory whose quantisation and counting of states is 
consistent with the results of Bekenstein and Hawking. 
Moreover, since  IH replaced the global notion of event horizons with a local 
description, the requirement of a 
knowledge of full space-time history as well as the asymptotics is avoided (see 
\cite{Corichi:2013zza, Corichi:2014zoa} for a first order description of 
theories with topological terms). The underlying spacetime therefore might not 
admit a 
global Killing vector at all in the isolated framework. While this has been a 
significant development in 
the understanding of black hole mechanics, generalizations to dynamically evolving  
horizons has
also been reported \cite{Ashtekar:2002ag, Ashtekar:2003hk, Ashtekar:2004cn}. These dynamical horizons are closely related to the notion of trapping horizons developed earlier \cite{Hayward:1993wb, Hayward:1994yy}. Using the boundary conditions for dynamical horizon it was shown that a flux balance law, relating the change of area of the 
dynamical horizon
to the flux of the matter energy, exists, reproducing an integrated version of a first law 
\cite{Ashtekar:2002ag, Ashtekar:2003hk, Ashtekar:2004cn}. Moreoever,
it has also been shown that if the horizon is slowly evolving, a form of the 
first law arises
\cite{Booth:2003ji, Booth:2006bn, Booth:2007wu}. The construction of a phase 
space
for these horizons has also been carried out in the metric variables. 

Another class of horizons that has been of interest are conformal Killing 
horizons (CKH). Though 
not a trapping horizon it essentially captures a dynamical situation. The 
notion of CKH
and it's properties were developed in \cite{DyerHonig,Suldyer, Sultana:2005tp, 
Jacobson:1993pf, Nielsen:2012xu}. These are null hypersurfaces whose null
geodesics are orbits of a conformal Killing field. If $\xi^{a}$ is a vector 
field which satisfies $\lie_{\xi}g_{ab}=2fg_{ab}$,
and is null, it generates a CKH for the metric $g_{ab}$. It has been shown that 
an analogue of the zeroth 
law holds for a conformal Killing horizon as well. More precisely, since $\xi^{a}$ generates a 
null surface, it is geodesic and one can define an accelration through
$\xi^{b}\nabla_{b}\xi^{a}=\kappa_{\xi} \xi^{a}$. Then, the quantity $(\kappa_{\xi}-2f)$ which 
essentially is a combination of the acceleration 
of the conformal Killing vector and the conformal factor, can be shown to be Lie dragged along 
the horizon and can therefore be 
interpreted as a temperature. An analogue of the first law is therefore 
expected to hold in this case as well but
has not been established in the literature. In this paper, we address the 
question if a form of the first law can be established 
at all for a CKH. As we discuss below, if such a law exists, it may lead to 
some important clues for a dynamically evolving horizon.

The plan of the paper is as follows. We start by developing the geometry of a 
quasi-local conformal Killing horizon. We assume that a 
spacetime time region $\mathcal{M}$ has a null boundary $\Delta$ which however 
may have non- zero expansion ($\theta =-2\rho\neq 0$). In other words we take 
the null generators of $\Delta$ to be only shear-free. We observe that these 
conditions are enough to ensure that the null generators $l^{a}$ are conformal 
Killing vectors on $\Delta$. Now, since these null surfaces are not expansion 
free, they may
be growing; in fact $\lie_l ~{}^{2}\epsilon=\theta~{}^{2}\epsilon$ and hence, are good 
candidates for growing
horizons. The situation in some sense mimics what one has at null infinity in 
an asymptotically flat space-time. 
However, we are more interested in an inner horizon. The physical situation for 
these horizons can be visualised
as follows. Suppose matter falls in through a horizon as a result of which it 
grows (supposing that matter satisfies standard energy conditions) and hence 
has a non- zero positive expansion. When this matter flux stops to fall in 
through the horizon, by the Raychaudhuri equation, an initially positively 
expanding horizon will slow down it's expansion and after some time reach the 
state of equilibrium. This equilibrium
state has zero expansion and it's geometrical set- up has been developed 
through the Isolated horizon formulation. We are interested to 
construct the space of solutions of only those dynamically evolving horizons 
which can be generated by a conformal Killing vector field. By construction, 
the CKH admit a limit to the IH formulation. We suppose that the matter flux 
across $\Delta$  be a real scalar field ($\varphi$)  satisfying the condition 
$\lie_l\varphi =-2\rho\varphi$ on the horizon. The geometrical conditions 
ensures that a form of zeroth law exists. In the next section, we show that the 
action for general relativity admits a well defined variational principle in 
presence of the conformal Killing horizon boundary and proceed to construct the 
symplectic structure. An interesting outcome is the 
construction of the phase space, identification of a boundary symplectic 
structure and the existence of a first law. Further, it arises 
that gravity and matter together gives a well defined phase- space provided a 
balance condition holds. This balance condition
turns out to be nothing but Einstein's equation contracted with the null 
generators $l^a$ (say). We thus get
a quasi-local analogue of a conformal Killing horizon.

\section{Geometrical setting and boundary conditions}
In this section, we introduce the minimal set of boundary conditions which 
are suitable for a quasilocal conformal Killing horizon. 
We assume that all fields under consideration are smooth. Let $\mathcal{M}$ be 
a $4$- manifold equipped
with a metric $g_{ab}$ of signature $(-,+,+,+)$. Consider a null hypersurface 
$\Delta$ of $\mathcal M$ with $l^a$ 
being it's future directed null normal. Given this null normal $l^{a}$, 
one can introduce another future directed null vector field $n^{a}$ which is 
transverse to $\Delta$. 
Further, one has a set of complex null vector field $(m,\,\bar{m})$, which are 
tangential to $\Delta$. 
This null tetrad $(l,\, n\, m\, \bar{m})$ constitutes the Newman- Penrose 
basis. The vector fields
satisfy the condition that $l.n=-1=-m.\bar{m}$, while all other scalar products 
vanish. Let $q_{ab}$ be the degenerate metric on the 
hypersurface. The expansion $\theta_{l}$ of the null normal is given by 
$q^{ab}\nabla_{a} l_b$. In terms of the Newman- Penrose co-effecients,
$\theta_l=-2\rho$ (see appendix A and \cite{Chandrasekhar:1985kt} for details). The accelaration of $l^a$ 
follows from the 
expression $l^a\nabla_a~l_b=(\epsilon+\bar\epsilon)l_b$ and is given by 
$\kappa_l :=\epsilon+\bar\epsilon$. To avoid cumbersome notation, we will do 
away with the subscripts $(l)$ from now on
if no confusion arises. It would be useful to define an equivalance class of 
null normals $[l^{a}]$ such that 
two null normals $l$ and $l'$ will be said to belong to the same equivalance 
class if $l'=cl$ where $c$ is a constant on $\Delta$.\\

\subsection*{Quasi-local conformal horizon}

{\it{Definition}}: A null hypersurface $\Delta$ of $\mathcal{M}$ will be called 
quasi-local conformal horizon if 
the following conditions hold.:
\begin{enumerate}
\item $\Delta$ is topologically  $S^2\times R$ and null.
\item The shear $\sigma$ of $l$ vanishes on $\Delta$ for any null normal $l$.
\item All equations of motion hold at $\Delta$ and the stress-
energy tensor $T_{ab}$ on $\Delta$ is such that $-T^{a}{}_{b}\,l^b$
is future directed and causal.
\item If $\varphi$ is a matter field then it must satisfy 
$\lie_l\varphi=-2\rho\,\varphi$ on $\Delta$ for all null normals $l$.
\item  The quantity $\left[2\rho+\epsilon + \bar{\epsilon}\right]$ is Lie 
dragged for any null-normal $l$.
\end{enumerate}

Some comments on the boundary conditions are in order. 
The first condition imposes restrictions on the topology of the hypersurface. 
It is natural to motivate this condition from Hawking's 
theorem on the topology of black holes in asymptotically flat stationary 
spacetimes or it's extension \cite{Hawking:1971vc, Galloway:2005mf}. But, we 
are also interested in spacetimes which are aymptotically 
non- flat or that they are non- stationary for which, these theorems may not 
hold true. However it is not unnatural to argue
that since black hole horizons forming out of gravitational collapse have 
spherical topologies, such conditions might exist. This condition is also  assumed in the Isolated Horizon formalism. 
For these isolated hypersurfaces,
the expansion $\theta$ of the null normal $l^{a}$ vanishes (which is not true 
in our case).
It is possible that cross- sections of such quasilocal horizons may admit other 
topologies. For the time being, we 
would not include such generalities and only retain the condition that the 
cross- sections of the hypersurfaces are spherical.

The second boundary condition on the shear is a simplification. Shear measures 
the amount of gravitational flux 
flowing across the surface, and we put the gravity flux to be vanishing. This 
boundary condition on 
the shear $\sigma$ of null normal $l^{a}$ has several consequences. First, 
since $l_a$ is hypersurface orthogonal, the Frobenius theorem implies that 
$\rho$ is real and $\kappa=0$. 
Secondly, the Ricci identity can be written as 
\beq
D\sigma-\delta\kappa=\sigma(\rho+\bar{\rho}+3\epsilon-\bar{\epsilon}
)-\kappa(\tau-\bar{\pi}+\bar{\alpha}+3\beta)+\Psi_0,
\eeq
where $D=l^{a}\nabla_{a}$, $\delta=m^{a}\nabla_{a}$, $\Psi_{0}$
is one of the Weyl scalars and the other quantities are the Newman- Penrose 
scalars (see \cite{Chandrasekhar:1985kt}
for details). If $\sigma\stackrel{\Delta}{=}0$, it implies 
$\Psi_0\stackrel{\Delta}{=}0$.
Next, since $\l^{a}$ is null normal to $\Delta$, it is twist- free and a 
geodetic vector field.
The implications of $\l^{a}$ being twist- free has already been shown above. 
The accelaration of $l^a$ follows from the 
expression $l^a\nabla_a~l^b=(\epsilon+\bar\epsilon)l^b$ and is given by 
$\kappa_l :=\epsilon+\bar\epsilon$. 
The acceleration of the null normal varies over the equivalence class $[cl]$ 
where $c$ is a constant on $
\Delta$ . This is only natural that
the acceleration varies in the class since in the absence of the knowledge of 
asymptotics, the acceleration cannot be fixed.

Further, it can be seen that the null normal $l^{a}$ is such that
\beq
\underleftarrow{\nabla_{(a}l_{b)}}\stackrel{\Delta}{=}-2\rho\, 
m_{(a}\bar{m}_{b)}
\eeq
which implies that $l^a$ is a conformal Killing vector on $\Delta$. Moreover, 
the Raychaudhuri equation implies that
$R_{ab}l^{a}l^{b}\neq 0$ and hence $-R^{a}{}_{b}l^{b}$ can have components 
which are tangential as well as transverse to $\Delta$.

The third boundary condition only implies that the field equations of gravity 
be satisfied and that the matter fields be such that 
their energy momentum tensor satisfies some mild energy conditions. The fourth 
and the fifth boundary conditions are somewhat adhoc 
but can be motivated. Let us first look at the fourth boundary condition. We 
have kept open the possibility that matter fields may
cross the horizon and the horizon may grow. The matter field is taken to be a 
massless scalar field $\varphi$ which behaves in a certain way
which mimics it's conformal nature. The fifth condition is motivated by the fact that
surface gravity remains invariant
under conformal transformations \cite{Jacobson:1993pf, Sultana:2005tp}. It can 
be shown that the quantity
that is constant for these horizons is 
$\left(2\rho+\epsilon+\bar{\epsilon}\right)$. A conformal 
transformation of the metric amounts to a conformal transformation of the 
two-metric on $\Delta$.
Under a conformal transformation $g_{ab}\rightarrow \Omega^2\,g_{ab}$ and one needs a new covariant derivative operator
which annihilates the conformally transformed metric. Under such a conformal transformation $l^a\rightarrow l^a,l_a\rightarrow \Omega^2 l_a,n^a\rightarrow \Omega^{-2} n^a,n_a\rightarrow n_a,m^a\rightarrow \Omega^{-1}m^a,m_a\rightarrow \Omega m_a$. The new derivative 
operator is such that it transforms as
\begin{equation}
\nabla_al_b\rightarrow\Omega^2\nabla_al_b+2\Omega\partial_a\Omega~l_b-\Omega^2\left[l_c\delta^c_a\partial_b\log{\Omega}+l_c
\delta^c_b\partial_a\log{\Omega}-g_{ab}g^{cd}l_c\partial_d\log{\Omega}\right]
\end{equation}
If one defines a one- form $\omega_a\=-n^b\underleftarrow{\nabla_a}l_b$, it 
transforms under the conformal transformation
as
\begin{equation}
\tilde{\omega}_a\=\omega_a+2\partial_a\log{\Omega}-\partial_a\log{\Omega}-n_al^c\partial_c\log{\Omega}
\end{equation}
It follows that the Newman- Penrose scalars transform in the following way
\begin{eqnarray}
\widetilde{\left(\epsilon+\bar\epsilon\right)}&\=&(\epsilon+\bar\epsilon) 
+2\lie_l\log{\Omega}\\
\tilde{\rho}&\=&\rho-\lie_l\log{\Omega}\\
\tilde{\sigma}&\=&\sigma
\end{eqnarray}
where, $\rho=-m^a\bar{m}^b\nabla_al_b$ and 
$\sigma=-\bar{m}^a\bar{m}^b\nabla_al_b$. Thus it follows that 
$2\rho + \epsilon+\bar\epsilon$ remains invariant under a conformal 
transformation.

At this point, it would be useful to recall the boundary conditions
of a weakly isolated horizon and note the important differences. A weakly 
isolated horizon
is  a null hypersurface which satisfies the first and the third boundary 
conditions given here
and that the expansion of the null normal $l^{a}$ be zero. On such surface, 
there 
exists a one- form $\omega_{a}$ which is also assumed to be Lie dragged by the 
vector field $l^{a}$.
Thus, instead of the condition on shear, for a WEH, the expansion of
the null normal $l^{a}$ is taken to be vanishing, $\theta=0=2\rho$. By the 
Raychaudhuri equation, the boundary conditions imply that
the shear is zero and that no matter field crosses the horizon (and hence the 
name isolated). However, here,
we impose only the condition that the shear vanishes and keep the possibility 
that matter fields may fall 
through the surface (but no gravitational flux) and that the hypersurface may 
grow along the affine parameter. 
As we shall show, removing our last condition does not restrict one to define a 
well defined phase space, but is
essential to get a first law. It is an analogue of the condition 
$\lie_l(\epsilon+\bar\epsilon)\stackrel{\Delta}{=}0$ assumed 
in the case of weakly isolated horizon. It may be useful to note that
the fifth boundary condition as given above, can be recast is a form which 
is an analogue of that for a weakly isolated horizon
by setting $\lie_l\tilde{\omega}=0$, where 
$\tilde{\omega}_a\=\omega_a+\partial_a\log{\Omega}-n_al^c\partial_c\log{\Omega}$
and the conformal factor is set such that $\lie_l\log{\Omega}=\rho$. 
\\

\section{Action principle and the classical phase space}
We are interested in constructing the space of solutions of general relativity, 
and we
use the first order formalism in terms of tetrads and connections. 
This formalism is naturally adapted to the nature of the problem in the sense 
that
the boundary conditions are easier to implement. Moreover it has the advantage 
that the 
construction of the covariant phase- space becomes simpler.
For the first order theory, we take the fields on the manifold to be
($e_{a}{}^{I},\, A_{aI}{}^{J},\, \varphi$), where $e_{a}{}^{I}$ is the co- 
tetrad, $A_{aI}{}^{J}$
is the gravitational connection and $\varphi$ is the scalar field.
The Palatini action in first order gravity with a scalar field is given by:
\begin{equation}\label{lagrangian1}
S_{G+M}=-\frac{1}{16\pi G}\int_{\mathcal{M}}\left(\Sigma^{I\!J}\wedge 
F_{I\!J}\right)-\frac{1}{2}\int_{\mathcal{M}}d\varphi\wedge {}{\star} 
d\varphi\;
\end{equation}
where $\Sigma^{IJ}=\half\,\epsilon^{IJ}{}_{KL}e^K\wedge e^L$, $A_{IJ}$ is a 
Lorentz $SO(3,1)$ connection 
and $F_{IJ}$ is a curvature two-form corresponding to the connection given by
$F_{IJ}=dA_{IJ}+A_{IK}\wedge A^{K}~_{J}$. The action might have to be 
supplemented with boundary terms to 
make the variation well defined.
\subsection*{Variation of the action}
For the variational principle, we consider the spacetime to be bounded by a 
null surface $\Delta$,
two Cauchy surfaces $M_{+}$ and $M_{-}$ which extend to the asymptotic 
infinity. The boundary conditions on the fields
are the following. At the asymptotic infinity, the fields satisfy appropriate 
boundary conditions. The 
fields on the hypersurfaces $M_{+}$ and $M_{-}$ are fixed so that their 
variations vanish. On the surface $\Delta$,
we fix a set of internal null- tetrad $(l^{I}, n^{I}, m^{I}, \bar{m}^{I})$ such 
that the flat
connection annihilates them. The fields on the manifold
($e_{a}{}^{I},\, A_{aI}{}^{J},\, \varphi$), must satisfy the following 
conditions. 
First, on $\Delta$, the configurations of the tetrads be such that 
$l^{a}=e_{I}^{a}l^{I}$
are the null vectors which satisfy the boundary conditions for quasi- local 
conformal horizon. 
Second, the possible connnections also satisfy the boundary conditions and be 
such that $\left(2\rho+\epsilon + \bar{\epsilon}\right)$ is constant.
Thirdly, we consider all those configurations of scalar field which, on 
$\Delta$, satisfy
$\lie_l\,\varphi=-2\rho\,\varphi$. 

We now check that the variational principle is well- defined if the boundary 
conditions on the fields, as given above, hold.
However, we need some expressions for tetrads and connections on $\Delta$, 
details of which are given in the appendix A.
On the conformal horizon, the $\Sigma^{IJ}$ is given by
\begin{equation} 
\underleftarrow{\Sigma}^{IJ}\=2l^{[I}n^{J]}~^{2}\epsilon+2n\wedge(im~l^{[I}\bar{
m}^{J]}-i\bar{m}~l^{[I}m^{J]}),
\end{equation}
and the connection is given by 
\begin{eqnarray}\label{connection_delta}
\underleftarrow{A_{a}{}_{IJ}}&\stackrel{\Delta}{=}&
2\left[(\epsilon+\bar{\epsilon})n_a 
-(\bar{\alpha}+\beta)\bar{m}_a-(\alpha+\bar{\beta})m_a\right]\, 
l_{[I}n_{J]}+2(-\bar{\kappa}n_a 
+\bar{\rho}\bar{m}_a)\, m_{[I}n_{J]}+2(-{\kappa}n_a 
+{\rho}{m}_a)\,\bar{m}_{[I}n_{J]}\nn
&+& 2(\pi n_a+-\mu\bar{m}_a-\lambda m_a)\, m_{[I}l_{J]}+2(\bar{\pi} n_a 
-\bar{\mu}{m}_a-\bar{\lambda}\bar{m}_a)\,\bar{m}_{[I}l_{J]}\nn
&+& 2\left[-(\epsilon-\bar{\epsilon})n_a +(\alpha-\bar{\beta}) m_a+(\beta-\bar{\alpha})\bar{m}_a
 \right]\, m_{[I}\bar{m}_{J]}..
\end{eqnarray}

The Lagrangian $4$- form for the fields ($e_{a}{}^{I},\, A_{aI}{}^{J},\, 
\varphi$) is given in the following way.
\begin{equation}
L_{G+M}=-\frac{1}{16\pi G}\left(\Sigma^{I\!J}\wedge 
F_{I\!J}\right)-\frac{1}{2}d\varphi\wedge \star d\varphi .
\end{equation}
The first variation of the action leads to equations of motion and boundary 
terms. The equations of motion consist of the following equations. 
First, variation of the action with respect to the connection implies that the 
curvature $F^{IJ}$ is related to the 
Riemann tensor $R^{cd}$,  through the relation 
$F_{ab}{}^{IJ}=R_{ab}{}^{cd}\,e^{I}_{c}e^{J}_{d}$. Second, variation with 
respect 
to the tetrads lead to the Einstein equations and third, the first variation of 
the matter 
field gives the equation of motion of the matter field.  
On- shell, the first variation is given by the following boundary terms
\begin{equation}
\delta L_{G+M} := d\Theta(\delta)=-\frac{1}{16\pi 
G}d\left(\Sigma^{IJ}\wedge\delta A_{IJ}\right)-d(\delta\varphi\star d\varphi),
\end{equation}
which are to be evaluated on the boundaries $M_{-}$, $M_{+}$, asymptotic 
infinity and $\Delta$. However, since fields are 
set fixed on the initial and the final hypersurfaces they vanish. The boundary 
conditions at infinity are
assumed to be appropriately chosen and they can be suitably taken care of. The 
only terms which are of relevance for this case are the terms on the internal boundary. 
On the internal boundary $\Delta$, the boundary terms give (see appendix 
\ref{appb} for details) 
\beq
16\pi G\, \delta L_{G+M}=-\delta\left(\frac{{\bf 
R}_{11}}{\rho}~n\right)\wedge{}^2\epsilon
-\delta\left(2\rho\, n\wedge\,^2\epsilon\right)
+ 8\pi G\,\delta\left(\frac{{\bf T}_{11}}{\rho}~n\right)\wedge~^2\epsilon
\eeq
Since Einstein's equations give $R_{11}=8\pi G \, T_{11}$, the first and the 
third term cancel and only
$\left(2\rho\, n\wedge\,^2\epsilon\right)$ remains. Thus, if one adds the
term $16\pi G\,S^{'}= \int_{\Delta}\left(2\rho\, n\wedge\,^2\epsilon\right)$ to the action, 
it is well defined
for the set of boundary conditions on $\Delta$. As we shall see below, since 
this is a boundary term,
it does not contribute to the symplectic structure.

\subsection*{Covariant phase- space and the symplectic Structure}

For a general Lagrangian, the on-shell variation gives $\delta 
L=d\Theta(\delta)$ 
where $\Theta$ 
is called the symplectic potential. It is a $3$-form in space-time and a 
$0$-form in phase space.
Given the symplectic potential, one can construct the symplectic structure 
$\Omega (\delta_{1}, \,\delta_{2})$ on the space of
solutions. One first constructs the symplectic current $J(\delta_1,\delta_2)= 
\delta_1\Theta(\delta_2)-\delta_2\Theta(\delta_1)$, 
which, by definition, is closed on-shell. The symplectic structure is then 
defined to be:
\begin{equation}
\Omega(\delta_1,\delta_2)=\int_{M}J(\delta_1,\delta_2)
\end{equation}
where $M$ is a space-like hypersurface. It follows that $dJ=0$
provided the equations of motion and linearized equations of motion hold. This 
implies that 
when integrated over a closed region of spacetime bounded by $M_+\cup M_-\cup 
\Delta$ (where $\Delta$  
is the inner boundary considered),
\begin{equation}
\int_{M_+}J-\int_{M_-}J~+~\int_{\Delta}J=0,
\end{equation}
where $M_+,M_-$ are the initial and the final space-like slices, respectively.
If the third term vanishes then the bulk symplectic structure is independent of 
choice of hypersurface. However, if 
it does not vanish but turns out to be exact, $\int_{\Delta}J=\int_{\Delta}dj $
then the hypersurface independent symplectic structure is given by:
\begin{equation}
\Omega(\delta_{1}, \,\delta_{2})= \int_MJ-\int_{S_\Delta}j
\end{equation}
where $S_\Delta$ is the 2-surface at the intersection of the hypersurface  $M$ with 
the boundary $\Delta$. 
The quantity $j(\delta_1,\delta_2)$ is called the boundary symplectic current 
and symplectic structure is
also independent of the choice of hypersurface.

Our strategy shall be to construct the symplectic structure for the action 
given in eqn. \eqref{lagrangian1}. Let us first look 
at the Lagrangian for gravity. The symplectic potential in this case is 
given by, $16\pi G\Theta(\delta)=-\Sigma^{I\!J}\wedge \delta A_{I\!J}$. The 
symplectic current 
is therefore given by,
\begin{equation}\label{symplectic_current1}
J_G(\delta_1,\delta_2)=-\frac{1}{8\pi 
G}\,\delta_{[1}\Sigma^{IJ}\wedge~\delta_{2]}A_{IJ}
\end{equation}
The above expression eqn. \eqref{symplectic_current1}, when pulled back and 
rescticted to the surface
$\Delta$ gives
\beq\label{symplec_pulled_back}
\underleftarrow{\delta_{[1}\Sigma^{IJ}\wedge\delta_{2]}A_{IJ}}
&\stackrel{\Delta}{=}&-2\,\delta_{[1}~^2{\bf\epsilon}
\wedge\delta_{2]}\left\{(\epsilon+\bar{\epsilon})n-(\alpha+\bar{\beta}
)m-(\bar{\alpha}+\beta)\bar{m}\right\}\nn
&&+2\,\delta_{[1}(n\wedge im)\wedge\delta_{2]}(\bar{\rho}\bar{m})
-2\,\delta_{[1}(n\wedge i\bar{m})\wedge\delta_{2]}(\rho m)
\eeq
It can be shown that the symplectic current pulled back on to $\Delta$ for the gravity sector is given by 
(see the appendix for details)\footnote{The entire construction and whatever follows goes through for negative $\rho$ with the replacement $\lvert\rho\rvert$ in place of $\rho$ in the argument of $\log$}
\beq
\\
\underleftarrow{J_G}(\delta_1,\delta_2)&\stackrel{\Delta}{=}&-\frac{1}{4\pi G}
\left[d\left(\delta_{[1}~^2{\bf\epsilon}~\delta_{2]}\log{\rho}\right)+\delta_{[1
}~^2{\bf\epsilon}\wedge\delta_{2]}
\left\{\left(\frac{\Phi_{00}}{\rho}
\right)n\right\}\right]
\eeq
The first term in the above expression is exact but not others. Therefore the 
phase is well 
defined for our boundary conditions $\sigma\stackrel{\Delta}{=}0$ provided,
if either $\Phi_{00}=0$, there is no matter flux across the horizon or if
$\Phi_{00}/\rho$ gets cancelled with a contribution from the matter degrees of 
freedom through Einstein's equation.
We deal with a 
more general case. We show that the
contribution of the scalar field is such that the symplectic current on $\Delta$ is again 
exact.

The symplectic current for the real scalar field is given by, 
$J_M(\delta_1,\delta_2)=2\,\delta_{[1}\varphi~\delta_{2]}\,{}\star d\varphi$.
The symplectic current on the hypersurface $\Delta$ can be obtained as
\begin{equation}
\underleftarrow{J_M}(\delta_1,\delta_2)=2\delta_{[1}\varphi~\delta_{2]}
(D\varphi~ n\wedge im\wedge\bar{m}),
\end{equation}
where $D=l^{a}\nabla_{a}$. The boundary condition on the scalar field implies 
$D\varphi=-2\rho \,\varphi$ and hence, we get 
that 
\begin{eqnarray}
\underleftarrow{J_M}(\delta_1,\delta_2)&=&4\delta_{[1}\varphi~\delta_{2]}
(-\varphi\,\rho~n\wedge im\wedge\bar{m})\\
&=&-d\left\{\delta_{[1}\varphi^2~\delta_{2]}~^2\epsilon\right\}+\delta_{[1}\frac
{D\varphi D\varphi}{\rho}n\wedge~\delta_{2]}
~^2\epsilon\nn
&=&-d\left\{\delta_{[1}\varphi^2~\delta_{2]}~^2\epsilon\right\}+\delta_{
[1}~^2\epsilon\wedge~\delta_{2]}
\left(\frac{{\bf T}_{11}}{\rho}n\right)
\end{eqnarray}
The combined expression is then given by:
\begin{equation}
\underleftarrow{J_{M+G}}(\delta_1,\delta_2) \stackrel{\Delta}{=}-\frac{1}{4\pi 
G}\left\{d\left(\delta_{[1}~^2{\bf\epsilon}~
\delta_{2]}\log{\rho}\right)\right\}-d\,\left\{\delta_{[1}\varphi^2~\delta_{2]}
~^2\epsilon\right\}
\end{equation}
It follows that the hypersurface independent symplectic structure is given by:
\begin{eqnarray}
\Omega(\delta_{1}, 
\delta_{2})=\int_{\mathcal{M}}J_{M+G}(\delta_1,\delta_2)-\int_{S_\Delta}
j&=&-\frac{1}{8\pi G}
\int_{\mathcal{M}}\delta_{[1}\Sigma^{IJ}\wedge~\delta_{2]}A_{IJ}+2\int_{\mathcal
{M}}\delta_{[1}\varphi~\delta_{2]}
(\star d\varphi)\nn
&+&\frac{1}{4\pi 
G}\int_{S_{\Delta}}\left\{\delta_{[1}~^2{\bf\epsilon}~\delta_{2]}\log{\rho}
\right\}
+\int_{S_{\Delta}}\delta_{[1}\varphi^2~\delta_{2]}~^2\epsilon
\end{eqnarray}
In the next section, we shall use this expression to derive the first law of 
mechanics for the conformal Killing horizon. 

\subsection*{Hamiltonian evolution and the first law}
Given the symplectic structure, we can proceed to study the evolution of the 
system. We assume that there exists a vector
which gives the time evolution on the spacetime. Given this vector field, one 
can define a corresponding vector field on the phase- space 
which can be interpreted as the infinitesimal generator of time evolution in 
the covariant phase- space. The Hamiltonian $H_l$ generating the 
time evolution is obtained as $\delta\, \tilde{H}_{l}= \Omega(\delta, \delta_{l})$, for 
all vector fields $\delta$ on the phase- space. Using
the Einstein equations, we get that
\begin{eqnarray}\label{firstlaw2}
\Omega(\delta,\delta_l)&=&-\frac{1}{16\pi 
G}\int_{S_{\Delta}}\left[(l.A_{IJ})\delta\Sigma^{IJ}-(l.\Sigma^{IJ})\wedge 
\delta A_{IJ}\right]
+\int_{S_{\Delta}}\delta\varphi~(l\cdotp{}{\star} d\varphi)\nn
\nn
&&\hspace{1cm}+\frac{1}{8\pi 
G}\int_{S_{\Delta}}\left(\delta~^2{\bf\epsilon}~\delta_{l}\log{\rho}-\delta_l~^2
{\bf\epsilon}
~\delta\log{\rho}\right)+\int_{S_{\Delta}}\frac{1}{2}(\delta\varphi^2~\delta_{l}
~^2\epsilon-\delta_l\varphi^2\delta~^2\epsilon)\nn
\end{eqnarray}

We now need to impose a few conditions on the fields to make a well defined 
Hamiltonian. These conditions are to be imposed since 
the action of $\delta_{l}$ on some phase- space fields is not like $\lie_{l}$. This is because of $\rho, \epsilon+\bar\epsilon$ and $\varphi$ all cannot be free data on $\Delta$.
First, we note the following equalities
\begin{eqnarray}
\lie_{l}\left(\frac{1}{4\pi G}\log{\rho}-\frac{1}{8\pi 
G}\log{\varphi}-\varphi^2\right)&=& \frac{1}{4\pi 
G}(2\rho+\epsilon+\bar{\epsilon})\\
\lie_{l}\left(\frac{~^2\epsilon}{\varphi}\right)&=&0
\end{eqnarray}
We assume that $\delta_l$ acts on $(2\rho+\epsilon+\bar{\epsilon})$ and 
$\left(\frac{~^2\epsilon}{\varphi}\right)$ like $\lie_l$. 
This can also be argued in the following fashion. Since $\delta_{l}\lie_{l}(2\rho+\epsilon+\bar{\epsilon})=0$ it 
immediately implies that 
$\lie_{l}\delta_{l}(2\rho+\epsilon+\bar{\epsilon})=0$. Hence, choosing
 $\delta_{l}(2\rho+\epsilon+\bar{\epsilon})=0$ at the initial cross-section implies that it remains zero throughout $\Delta$. Furthermore if we set $\delta_l\left(\frac{1}{4\pi G}\log{\rho}-\frac{1}{8\pi 
G}\log{\varphi}-\varphi^2\right)$=0 
at the initial cross-section, it remains zero everywhere on $\Delta$ and so,
\begin{eqnarray}\label{eqn_no1}
\frac{\delta_l\rho}{\rho}-8\pi 
G\varphi\delta_l\varphi-\frac{\delta_l\varphi}{2\varphi}=0
\end{eqnarray}
Another condition can be derived from the equation above
\begin{eqnarray}
\delta_l\left(\frac{~^2\epsilon}{\varphi}\right)=\frac{1}{\varphi}
\delta_l~^2\epsilon-~^2\epsilon\frac{1}{\varphi^2}\delta_l\varphi=0
\end{eqnarray}
The variations $\delta_l$ satisfy the following differential equations, which 
can be checked to be consistent with each other:
\begin{eqnarray}
\label{eqn_no2}
\lie_l\delta_l\varphi&=&-2\delta_l\rho\varphi-2\rho\delta_l\varphi\\
\lie_l\delta_l~^2\epsilon&=&-2\delta_l\rho~^2\epsilon-2\rho\delta_l~^2\epsilon
\end{eqnarray}
Putting condition $\eqref{eqn_no1}$ in $\eqref{eqn_no2}$, we get
\begin{equation}
\delta_l\varphi=C(\theta,\phi)\exp\left[-{\int\left(16\pi 
G\varphi^2+3\right)\rho dv}\right],
\end{equation}
where $C(\theta,\phi)$, is a constant of integration. If we choose this 
constant $C(\theta,\phi)=0$, it immediately implies that 
$\delta_{l}\varphi=0=\delta_{l}{}^{2}\epsilon.$ With the choice of $\delta_l$ 
only the bulk symplectic structure survives and one gets from eq. $\eqref{firstlaw2}$ \footnote{We assume that the contribution from the boundary at asymptotic infinity is a total variation $\delta E^\infty$.}
\begin{eqnarray}
\delta H_{l}&=&-\frac{1}{8\pi 
G}\int_{S_\Delta}(2\rho+\epsilon+\bar{\epsilon})\delta~^2\epsilon+
\frac{1}{8\pi G}\int_{S_\Delta}~^2\epsilon~(-\delta\rho-8\pi G\,\delta\varphi 
D\varphi)+\delta E^\infty
\end{eqnarray}
where we have redefined our Hamiltonian $H_l=\tilde{H}_l+\int_{S_\Delta}\rho~^2\epsilon$. This redefination is possible since the definition of the Hamiltonian is ambiguous upto a total variation. Further, as expected $\Omega(\delta_{l},\delta_l)=0$. Next we define, $E^l_{\Delta}=E^\infty-H_l,$ as the horizon energy. It is clear from above that for $\rho\rightarrow 0$ (i.e in the isolated horizon limit) it matches with the definition in \cite{Ashtekar:2002ag,Ashtekar:2003hk} if asymptotics is flat and $E^\infty=E_{ADM}$. It therefore follows that:
\beq
-{\delta}E^l_\Delta =-\frac{1}{8\pi G}\int_{S_\Delta}(2\rho+\epsilon+
\bar{\epsilon}){\delta}~^2\epsilon-\frac{1}{8\pi 
G}\int_{S_\Delta}\left[~^2\epsilon~({\delta}\rho+8\pi G\,{\delta}\varphi 
D\varphi)\right].
\eeq

To recover the the 
more familiar form of first law known for a dynamical situation, we  assume 
there is a vector field $\tilde{\delta}$ on phase space which acts only on the 
fields on
$\Delta$ (and not in the bulk) such that it's action on the boundary variables 
is to evolve the boundary 
fields along the affine parameter $v$ (it may be interpreted to be a time 
evolution, like $\lie$). Now demanding that $\tilde{\delta}$ to be 
Hamiltonian would give an integrability condition which also ensures that 
$\delta_l$ is Hamiltonian. So 
one can calculate $\Omega(\tilde{\delta},\delta_l):=\tilde{\delta}H_{l}$
%
%
which can be written in the following form\footnote{If the stress tensor satisfies the dominant enegy condition then $(2\rho+\epsilon+\bar\epsilon)$ is a constant on $\Delta$ \cite{Sultana:2005tp}.}
\begin{equation}\label{firstlaw}
\dot{E^l_\Delta}=\,\frac{1}{8\pi G}\left(2\rho+\epsilon+\bar{\epsilon}\right)\dot{A}+\frac{1}{8\pi 
G}\int_{S_\Delta}\left[~^2\epsilon~(\dot\rho+8\pi 
G\,\dot\varphi D\varphi)\right]
\end{equation}
where dots imply changes in the variables produced by the action of $\tilde{\delta}$.
%
%
Note that if $\tilde{\delta} =\lie_{l}$, then, $\tilde{\delta}\varphi D\varphi$ gives the expression $T_{ab}l^{a}l^{b}$.
Equation \eqref{firstlaw} is the form of evolution for the conformal Killing horizons. The first term in the above expression is the usual $TdS$ term while the second term is a flux term which takes into account the non-zero matter flux across $\Delta$.

\section{Discussions}

In this paper, we have developed the geometrical set-up for a quasi-local 
description
of a conformal Killing horizon. Further, we have also shown that one can 
understand these horizons
to have a zeroth law (as was also discussed in \cite{Suldyer}) and a first law. This 
development of a notion of quasi-local conformal horizon should be taken in the 
same spirit as the development of the notion of isolated horizon from Killing 
horizons. A conformal Killing horizon is one which has a conformal Killing 
vector in the neighbourhood of the horizon.  In contrast, a quasi-local conformal 
horizon only requires the existence of a null hypersurface generating vector 
which is shear free on the null hypersurface. 
The number of solutions of Einsteins's equation for gravity and matter that admits 
a conformal Killing horizon may be 
small (examples of such kind has been constructed by \cite{Sultana:2005tp}). 
However the solutions admitting a 
quasi-local conformal horizon may be large.  We do not comment on the nature of 
solutions that admits a quasi-local conformal horizon,
we think that significant amount of insights may be obtained by numerical 
simulations and therefore falls in the regime of numerical relativity. 
The most useful application of these geometrical structures are in the 
dynamical evolution of black holes. Indeed,
as matter falls in through the horizon and the black hole horizon grows, the 
expansion is non- zero. In such cases, it is important to 
understand if in this dynamical situation 
one can prove the existence of laws for black hole mechanics in 
some form.

We have taken a real scalar field as the  matter field in question. The flux 
balance law is seen to be 
successfully implemented if it satisfies a condition 
$\lie_l\varphi\stackrel{\Delta}{=}-2\rho\,\varphi$. This 
assumption is motivated through the fact that $l^a$ is a conformal Killing 
vector on $\Delta$. Taking 
other matter fields will therefore be an immediate extension of our work. 
Further, from the onset we have ignored any space-like axial
conformal Killing vector on $S_{\Delta}$. So a generalization to the
rotational case seems to be another plausible extension. Since the case of an 
isolated horizon appears as a special case $\rho\rightarrow0$, the consistency 
of our analysis can actually be checked by taking the isolated horizon limit. 
In fact we perform this consistency check and find that 
the final expressions and the first law
does give back the results obtained for an isolated horizon. 

We should mention at this point that our construction does not capture the most general dynamical situation, as constructed in \cite{Hayward:1993wb, Ashtekar:2002ag}. The horizons discussed in these references are spacelike boundaries foliated by partially trapped two surfaces which may not be shear-free. Further, an integrated version of the first law has been demonstrated to exist, which captures the dynamics of growing black hole horizons in full generality. However in these constructions, which use metric variables, the existence of a well defined phase- space has not been established and consequently the first law does not follow directly from the symplectic structure. In our case we have assumed that there is no gravitational flux (shear is zero) but only matter flows across the null boundary $\Delta$. In this simplified geometry, we have demonstrated that a space of solutions of Einstein's equations exists which admit the boundary conditions of CKH and that a differential version of the first law of black hole mechanics can be obtained. Also, we have used the first order formalism for the construction of this symplectic structure. We do not know if one may get a well defined symplectic structure for boundary conditions discussed in \cite{Hayward:1993wb, Ashtekar:2002ag}. Even if one is able to construct a phase- space, it is not possible to obtain a differential version the first law since there is no analogue of the zeroth law for such boundaries, but an integrated version of the first law is expected to hold.

Given  a form of the first law, it is obvious to compare with the first law of 
thermodynamics. However, since the horizon is growing, 
it describes a non- equilibrium situation and hence may differ considerably 
from equilibrium thermodynamics where 
one studies the transition from one equilibrium state to a nearby equilibrium 
state.  One should keep in view that thermodynamics arises
out of microscopic dynamics of the underlying degrees of freedom and have 
universal validity (that are independent
of the underlying dynamics of a particular system). For a general dynamical 
spacetime (when the gravitational degrees of freedom
are excited), there is no time translation symmetry and hence no definition of 
entropy may be possible. 
Also in non- equilibrium cases, a system may not get enough time to relax 
back to the equilibrium state and hence no canonical
definition of temperature exists. But, in the present scenario, though the 
horizon makes transition between two states which are far from equilibrium, 
because there exits a conformal Killing vector, this leads to a definite 
identification of temperature  and a first law and possibly entropy.
One may then enquire if dynamically growing horizons is attributed some 
entropy that can arise from some counting
of microstates. 
The boundary symplectic structure has a natural interpretation of being the 
symplectic structure of a field theory
residing on the boundary. In the case of an isolated horizon it turns out to be 
an $SU(2)$ or an $U(1)$ Chern-Simons
theory. A quantization of the boundary theory therefore provides a microscopic 
description of the entropy of the 
isolated horizon. Since we explicitly construct 
the boundary symplectic structure it will be interesting to see if it does 
coincide with any known topological field theory. 
A complete answer to such questions shall have important implications for 
thermodynamics as well as black hole physics.

\appendix

\section{The Connection in terms of Newman-Penrose co-effecients}
Fix a set a internal null vectors $(l_I,n_I,m_I,\bar m_I) $ on $\Delta$ such  
that $\partial_a (l_I,n_I,m_I,\bar m_I)\stackrel{\Delta}{=}0$. Given any tetrad 
$e^I_a$, the null tetrad $(l_a,n_a,m_a,\bar m_a)$ can be expanded as 
$l_a=e^I_a~l_I$. The expression for $\Sigma^{IJ}$ can now be readily calculated 
and is given as.
\beq
\Sigma^{IJ}&=&2l^{[I}n^{J]}~^{2}\epsilon+2n\wedge(im~l^{[I}\bar{m}^{J]}-i\bar{m}
~l^{[I}m^{J]})\nn
\nn
&-&2i~l\wedge 
n~m^{[I}\bar{m}^{J]}-2l\wedge(im~n^{[I}\bar{m}^{J]}-i\bar{m}~n^{[I}m^{J]})
\eeq

This is the full expression for $\Sigma^{IJ}$ where nothing has been been assumed 
regarding the nature of the
 boundary $\Delta$. If $\Delta$ is a null surface and $l_{a}$ is the null 
normal, we get that 
\beq
\underleftarrow{\Sigma}^{IJ}&\stackrel{\Delta}{=}&2l^{[I}n^{J]}~^{2}\epsilon+2n\wedge(im~l^{[I}\bar
{m}^{J]}-i\bar{m}~l^{[I}m^{J]})\nn
\eeq

The covariant derivative is defined to be compatible with the tetrad 
\emph{i.e.} $\nabla_b~e^I_a=0$. The covariant derivatives on the null tetrads 
can be written in terms of the Newman-Penrose coeffecients and are given by the 
following,

\beq
\nabla_al_b&=&-(\epsilon+\bar{\epsilon})n_al_b+\bar{\kappa}n_am_b+\kappa
n_a\bar{m}_b-(\gamma+\bar{\gamma})l_al_b+\bar{\tau}l_am_b+\tau l_a\bar{m}_b\nn
&&+[(\bar{\alpha}+\beta)\bar{m}_al_b
-\bar{\rho}\bar{m}_am_b-\sigma\bar{m}_a\bar{m}_b+(\alpha+\bar{\beta}
)m_al_b-\rho m_a\bar{m}_b-\bar{\sigma}m_a m_b]\nn
\\
\nabla_a n_b&=&(\epsilon+\bar{\epsilon})n_an_b-\pi 
n_am_b-\bar{\pi}n_a\bar{m}_b+(\gamma+\bar{\gamma})l_an_b-\nu l_am_b-\bar{\nu} 
l_a\bar{m}_b\nn
&&-[(\bar{\alpha}+\beta)\bar{m}_an_b-\mu\bar{m}_am_b-\bar{\lambda}\bar{m}_a\bar{
m}_b+(\alpha+\bar{\beta})m_an_b-\bar{\mu} m_a\bar{m}_b-\lambda m_a m_b]\nn
\\
\nabla_am_b&=&-\bar{\pi}n_al_b+\kappa
n_an_b-(\epsilon-\bar{\epsilon})n_am_b-\bar{\nu}l_al_b+\tau 
l_an_b-(\gamma-\bar{\gamma})l_am_b\nn
&&+[\bar{\lambda}\bar{m}_al_b-\sigma\bar{m}_an_b+(\beta-\bar{\alpha})\bar{m}
_am_b+\bar{\mu} m_al_b-\rho m_an_b+(\alpha-\bar{\beta})m_a{m}_b]\nn
\eeq

Next, once we have fixed a set of null internal vectors on $\Delta$, the 
connection can be expanded in terms of these Newman- Penrose
coefficients. Note that $\nabla_a~l_I=\partial_a~l_I+A_{aI}^J~l_J$. Therefore 
on $\Delta$,  we have 
$e^b_I\nabla_al_b\stackrel{\Delta}{=}A_{aI}~^Jl_J$ and hence

\beq
A^{(l)}_{aI}~^Jl_J&\stackrel{\Delta}{=}&-(\epsilon+\bar{\epsilon})n_al_I+\bar{\kappa}n_am_I+\kappa
n_a\bar{m}_I-(\gamma+\bar{\gamma})l_al_I+\bar{\tau}l_am_I+\tau l_a\bar{m}_I\nn
&&+[(\bar{\alpha}+\beta)\bar{m}_al_I-\bar{\rho}\bar{m}_am_I-\sigma\bar{m}_a\bar{
m}_I+(\alpha+\bar{\beta})m_al_I-\rho m_a\bar{m}_I-\bar{\sigma}m_a m_I]\nn
\\
A^{(l)}_{aIJ}&\stackrel{\Delta}{=}&\left[(\epsilon+\bar{\epsilon})n_a 
+(\gamma+\bar{\gamma})l_a-(\bar{\alpha}+\beta)\bar{m}_a-(\alpha+\bar{\beta}
)m_a\right]\, 2l_{[I}n_{J]}\nn
&&+\left[-\bar{\kappa}n_a-\bar{\tau}l_a+\bar{\rho}\bar{m}_a+\bar{\sigma}
m_a\right]\, 2m_{[I}n_{J]}+\left[-{\kappa}n_a-\tau 
l_a+{\rho}{m}_a+{\sigma}\bar{m}_a\right]\, 2\bar{m}_{[I}n_{J]}\nn
\eeq
where the subscript $l$ in $A^{(l)}$ indicates that the only the vector field 
$l^{a}$ has been used to evaluate the connection. Similarly, we can proceed for 
other vector fields $n^{a}, m^{a}$ and $\bar{m}^{a}$. The resulting connections 
are given as 
follows 
\beq
A^{(n)}_{aIJ}&\stackrel{\Delta}{=}&\left[-(\epsilon+\bar{\epsilon})n_a-(\gamma+\bar{\gamma})l_a+(\bar{\alpha}+\beta)\bar{m}_a+(\alpha+\bar{\beta})m_a\right]\, 2n_{[I}l_{J]}\nn&&+(\pi n_a+\nu l_a-\mu\bar{m}_a-\lambda m_a)\, 2m_{[I}l_{J]}+(\bar{\pi} n_a+\bar{\nu}l_a-\bar{\mu}{m}_a-\bar{\lambda}\bar{m}_a)\, 2\bar{m}_{[I}l_{J]}\nn
\\\nn A^{(m)}_{aIJ}&\stackrel{\Delta}{=}&(-\bar{\pi}n_a-\bar{\nu}l_a+\bar{\lambda}\bar{m}_a+\bar{\mu} m_a)\, 2l_{[I}\bar{m}_{J]}+(\kappa n_a+\tau l_a-\sigma\bar{m}_a-\rho m_a)\, 2n_{[I}\bar{m}_{J]}\nn&&+\left[-(\epsilon-\bar{\epsilon})n_a-(\gamma-\bar{\gamma})l_a+({\alpha}-\bar\beta){m}_a+(\beta-\bar{\alpha})\bar{m}_a\right]\, 2m_{[I}\bar{m}_{J]}
\eeq

The full connection is then given by:
\beq
A_{aIJ}&\stackrel{\Delta}{=}&2\left[(\epsilon+\bar{\epsilon})n_a+(\gamma+\bar{
\gamma})l_a-(\bar{\alpha}+\beta)\bar{m}_a-(\alpha+\bar{\beta} )m_a\right]\, 
l_{[I}n_{J]}\nn
\nn
&&+2\left[-\bar{\kappa}n_a-\bar{\tau}l_a+\bar{\rho}\bar{m}_a+\bar{\sigma}
m_a\right]\, m_{[I}n_{J]}+2\left[-{\kappa}n_a-\tau 
l_a+{\rho}{m}_a+{\sigma}\bar{m}_a\right]\, \bar{m}_{[I}n_{J]}\nn
\nn
&&+2\left[\pi n_a+\nu l_a-\mu\bar{m}_a-\lambda m_a\right]\, 
m_{[I}l_{J]}+2\left[\bar{\pi} 
n_a+\bar{\nu}l_a-\bar{\mu}{m}_a-\bar{\lambda}\bar{m}_a\right]\, 
\bar{m}_{[I}l_{J]}\nn
\nn
&&+2\left[-(\epsilon-\bar{\epsilon})n_a-(\gamma-\bar{\gamma})l_a+({\alpha}-\bar\beta){m}_a+(\beta-\bar{
\alpha})\bar{m}_a\right]\, m_{[I}\bar{m}_{J]}
\eeq

Note that as in in the case of $\Sigma_{IJ}$ no boundary condition has been 
assumed in the above expression. In the main part of the paper 
this expression for the connection eqn \eqref{connection_delta} shall be used 
but with the boundary conditions.

Further, we would be requiring the exterior derivatives on the null tetrads. We 
therefore give the expressions here. 
\beq
dn=\nabla_{a}n_{b}\, dx^a\wedge dx^b&=&-\pi n\wedge 
m-\bar{\pi}n\wedge\bar{m}+(\gamma+\bar{\gamma})l\wedge n-\nu l\wedge 
m-\bar{\nu} l\wedge\bar{m}\nn
&&\hspace{1cm}-[(\bar{\alpha}+\beta)\bar{m}\wedge n-\mu\bar{m}\wedge 
m+(\alpha+\bar{\beta})m\wedge n-\bar{\mu} m\wedge\bar{m}]\nn
\\
dl=\nabla_{a}l_{b}\, dx^a\wedge dx^b&=&-(\epsilon+\bar{\epsilon})n\wedge 
l+\bar{\kappa}n\wedge m+\kappa n\wedge\bar{m}+\bar{\tau}l\wedge m+\tau 
l\wedge\bar{m}\nn
&&\hspace{1cm}+[(\bar{\alpha}+\beta)\bar{m}\wedge l-\bar{\rho}\bar{m}\wedge 
m+(\alpha+\bar{\beta})m\wedge l-\rho m\wedge\bar{m}]\nn
\\
dm=\nabla_{a}m_{b}\, dx^a\wedge dx^b&=&-\bar{\pi}n\wedge 
l-(\epsilon-\bar{\epsilon})n\wedge m+\tau l\wedge 
n-(\gamma-\bar{\gamma})l\wedge m\nn
&&\hspace{1cm}+[\bar{\lambda}\bar{m}\wedge l-\sigma\bar{m}\wedge 
n+(\beta-\bar{\alpha})\bar{m}\wedge m+\bar{\mu} m\wedge l-\rho m\wedge n]
\eeq
From the above expressions, it follows that for the area two- form which is 
given by ${}^{2}\epsilon=im\wedge\bar{m}$, 
we get that  $d~^2\epsilon=2\rho \,n\wedge ~^2\epsilon \,\mbox{and}  \, \lie_l 
~^2\epsilon=-2\rho~~^2\epsilon$.

\section{Variation of the action}\label{appb}
Since the boundary symplectic structure turned out to be exact, it is at once 
evident that 
the variation of the action should be well-defined with the the boundary 
conditions considered. However one 
may need to add an additional boundary term in order to it. As has been pointed 
out that such terms won't affect
the symplectic structure though. Therefore for completeness we consider the 
variation of the action and find out the
necessary boundary term needed to make the variation well defined. We consider 
the action for gravity and a scalar field 
without any boundary terms a priori. The expression for $\Theta$ on $\Delta$ is 
calculated imposing the boundary conditions 
and the required boundary term can be obtained. We have 
\beq
L_{M+G}=-\frac{1}{16\pi G}\left(\Sigma^{I\!J}\wedge 
F_{I\!J}\right)-\frac{1}{2}d\varphi\wedge \star d\varphi;
\eeq
It follows that
\beq
d\Theta(\delta)=-\frac{1}{16\pi G}d\left(\Sigma^{IJ}\wedge\delta 
A_{IJ}\right)-d(\delta\varphi\star d\varphi)
\eeq
Consider the gravity terms first\footnote{In our case it might not be possible to define a unique covariant derivative on $\Delta$. However, since in the the calculations $l^a\nabla_a$ acts only on functions, the amibiguity do not play a role.}
\beq
\underleftarrow{\Sigma^{IJ}\wedge\delta A_{IJ}}&\stackrel{\Delta}{=}&-2~^2\epsilon\wedge\delta[(\epsilon+\bar{\epsilon})n]+2(n\wedge im)\wedge\delta(\rho\bar{m})-2(n\wedge 
i\bar{m})\wedge\delta(\rho{m})\nn
&=&-2~^2\epsilon\wedge\delta\left[\left(\frac{D\rho}{\rho}-\rho-\frac{\Phi_{00}}
{\rho}\right)n\right]+2(n\wedge im)\wedge\delta(\rho\bar{m})-2(n\wedge 
i\bar{m})\wedge\delta(\rho{m})\nn
&=&d\left[2~^2\epsilon~\delta(\log{\rho})\right]-4n\wedge 
{}^2\epsilon\,\delta\rho+2~^2\epsilon\wedge\delta\left[\left(\rho+\frac{\Phi_{00
}}{\rho}\right)n\right]+4n\wedge^2\epsilon~\delta\rho+2\rho 
n\wedge\delta~^2\epsilon\nn
\nn
&=&d\left[2~^2\epsilon~\delta(\log{\rho})\right]+2~^2\epsilon\wedge\delta\left[
\left(\frac{{\bf R}_{11}}{2\rho}\right)n\right]+\delta\left(2\rho 
n\wedge~^2\epsilon\right)
\eeq
The matter term gives

\beq
(\delta\varphi\star d\varphi)
&=&-d\left(\frac{1}{2}\delta\varphi^2~^2\epsilon\right)+\delta\left(\varphi~d\varphi\right)\wedge~^2\epsilon\nn
&=&-d\left(\frac{1}{2}\delta\varphi^2~^2\epsilon\right)-\frac{1}{2}\delta\left(\frac{{\bf T}_{11}}{\rho}n\right)\wedge~^2\epsilon
\eeq
Adding everything up, one finds that,
\beq
d\Theta(\delta)=-\frac{1}{16\pi G}d\left(\Sigma^{IJ}\wedge\delta 
A_{IJ}\right)-d(\delta\varphi\star d\varphi)=-\frac{1}{8\pi G}d\delta\left(\rho 
n\wedge~^2\epsilon\right)
\eeq
So one needs to add $\frac{1}{8\pi G}\int_{\Delta}\left(\rho n\wedge~^2\epsilon\right)$ to the action 
to make the variation well-defined.

\section{Boundary Symplectic Structure for Gravity}

The symplectic current in first order gravity is therefore given by,
\beq
J_G(\delta_1,\delta_2)&=&-\frac{1}{8\pi 
G}\delta_{[1}\Sigma^{IJ}\wedge~\delta_{2]}A_{IJ}
\eeq

We need to pull back the above expression on to the boundary and check if it is 
exact.

\beq
\underleftarrow{\delta_{[1}\Sigma^{IJ}\wedge\delta_{2]}A_{IJ}}&\stackrel{\Delta}{=}&-2\delta_{[1}~^2{
\bf\epsilon}
\wedge\delta_{2]}((\epsilon+\bar{\epsilon})n-(\alpha+\bar{\beta})m-(\bar{
\alpha}+\beta)\bar{m})\nn
&&+2\delta_{[1}(n\wedge im)\wedge\delta_{2]}(\bar{\rho}\bar{m})
-2\delta_{[1}(n\wedge i\bar{m})\wedge\delta_{2]}(\rho m)
\eeq

We consider the first term in the above expression. By using the Ricci identity 
in terms of Newman-Penrose co-effecients
\beq
D\rho=\rho^2+\rho(\epsilon+\bar{\epsilon})+\Phi_{00}
\eeq
we find that the first term can be written in the following form,
\beq
-2\delta_{[1}~^2{\bf\epsilon}\wedge\delta_{2]}((\epsilon+\bar{\epsilon})n)&=&-2\delta_{[1}~^2{\bf\epsilon}\wedge\delta_{2]}\left(\left(\frac{D\rho}{\rho}
-\frac{\rho^2}{\rho}-\frac{\Phi_{00}}{\rho}
\right)n\right)\nn
&&=d\left(2\delta_{[1}~^2{\bf\epsilon}~\delta_{2]}log{\rho}
\right)-\left(2\delta_{[1}~d^2{\bf\epsilon}\wedge\delta_{2]}log{\rho}\right)\nn
&&\hspace{4cm}+2\delta_{[1}~^2{\bf\epsilon}\wedge\delta_{2]}\left(\left(\frac{
\rho^2}{\rho}+\frac{\Phi_{00}}{\rho}\right)n\right)\nn
\eeq

Since the first term in the above expression is already exact, we leave it for 
the the
moment and check if there is any simplication of the other terms when combined 
with the rest of the 
third and forth term in the symplectic current.

\beq
-2\delta_{[1}~d^2{\bf\epsilon}\wedge\delta_{2]}log{\rho}
&&=-4\delta_{[1}~i\rho n\wedge m\wedge\bar{m}~\delta_{2]}log{\rho}\nn
&&=-2\delta_{[1}~(n\wedge im)\wedge\bar{m}~\delta_{2]}\rho-2(n\wedge 
im)\wedge\delta_{[1}\bar{m}~\delta_{2]}\rho\nn
&&\hspace{1cm}+2\delta_{[1}~(n\wedge i\bar{m})\wedge 
m~\delta_{2]}\rho+2(n\wedge i\bar{m})\wedge\delta_{[1} m~\delta_{2]}\rho
\eeq

The third and the fourth term in the symplectic current gives:
\beq
&&2\delta_{[1}(n\wedge im)\wedge\delta_{2]}(\rho\bar{m})
-2\delta_{[1}(n\wedge i\bar{m})\wedge\delta_{2]}(\rho m)\nn
&&\hspace{3cm}=2\delta_{[1}(n\wedge 
im)\wedge\bar{m}~\delta_{2]}(\rho)+2\rho\delta_{[1}(n\wedge 
im)\wedge\delta_{2]}\bar{m}\nn
&&\hspace{4cm}-2\delta_{[1}(n\wedge i\bar{m})\wedge 
m~\delta_{2]}(\rho)-2\rho\delta_{[1}(n\wedge i\bar{m})\wedge\delta_{2]}m
\eeq

Adding the above two equations and then simplifying gives:

\beq
&&-2\delta_{[1}~d^2{\bf\epsilon}\wedge\delta_{2]}log{\rho}+2\delta_{[1}(n\wedge 
im)\wedge\delta_{2]}(\rho\bar{m})
-2\delta_{[1}(n\wedge i\bar{m})\wedge\delta_{2]}(\rho m)\nn
&&\hspace{3cm}=-2n\wedge\delta_{[1}~^2\epsilon\delta_{2]}\rho+2\rho\delta_{[1}
(n)\wedge\delta_{2]}~^2\epsilon\nn
&&\hspace{3cm}=-2\delta_{[1}~^2\epsilon\wedge\delta_{2]}(\rho n)
\eeq

So the boundary term becomes
\beq
d\left(2\delta_{[1}~^2{\bf\epsilon}~\delta_{2]}log{\rho}\right)+2\delta_{[1}~^2{
\bf\epsilon}\wedge\delta_{2]}\left(\frac{
\Phi_{00}}{\rho}n\right)
\eeq

\section{Bulk Symplectic structure}
For any vector field $\xi$ generating diffeomorphisms, the corresponding phase 
space variation $\delta_\xi$ acts in the bulk like $\lie_\xi$. It can then be 
shown that
\beq
J_G(\delta,\delta_\xi)&=&-\frac{1}{16\pi 
G}\left[(\xi.A_{IJ})\delta\Sigma^{IJ}-(\xi.\Sigma^{IJ})\wedge \delta 
A_{IJ}\right]+(\text{Equations of motion})\delta e^I
\eeq
Similarly for the matter fields, we get that
\beq
J_M(\delta,\delta_{\xi})
&=&d\left[\delta\varphi~(\xi.\star d\varphi)\right]-\left[\delta 
d\varphi~(\xi.\star d\varphi)\right]-\xi.d\varphi~\delta(\star d\varphi)\\
\eeq
The second and the third term in the last expression enters Einstein's 
equation. Therefore the full bulk symplectic structure is,

\beq
\int_{M}J(\delta,\delta_\xi)&=&-\frac{1}{16\pi G}\int_{\partial 
M}\left[(\xi.A_{IJ})\delta\Sigma^{IJ}-(\xi.\Sigma^{IJ})\wedge \delta 
A_{IJ}\right]+\int_{\partial M}\delta\varphi~(\xi.\star d\varphi)
\eeq

\subsection*{Acknowedgements}
The authors acknowledge the discussions with Amit Ghosh. The authors also thank the anonymous referee for suggestions that made the presentation better. AC is partially supported through the UGC- BSR start-up grant vide their letter no. F.20-1(30)/2013(BSR)/3082. AG is supported by Department of Atomic-Energy, Govt. Of India.

\end{document}